\documentclass[fleqn,usenatbib]{mnras}

\usepackage{newtxtext,newtxmath}
\usepackage[T1]{fontenc}

\DeclareRobustCommand{\VAN}[3]{#2}
\let\VANthebibliography\thebibliography
\def\thebibliography{\DeclareRobustCommand{\VAN}[3]{##3}\VANthebibliography}

\usepackage{graphicx}	% Including figure files
\title[Search for an alien communication to Wolf 359]{Search for an alien communication from the Solar System to a neighbor star }

\author[M. Gillon, A. Burdanov, J. T. Wright]{
Micha\"el Gillon$^{1}$\thanks{E-mail: michael.gillon@uliege.be},
Artem Burdanov$^{2}$,
Jason T. Wrigth$^{3}$
\\
$^{1}$Astrobiology Research Unit, Universit\'e de Li\`ege, All\'ee du 6 Ao\^ut 19, B\^at. B5C, 4000 Li\`ege, Belgium\\
$^{2}$Department of Earth, Atmospheric and Planetary Science, Massachusetts Institute of Technology, 77 Massachusetts Avenue, Cambridge, MA 02139, USA\\
$^{3}$Penn State Extraterrestrial Intelligence Center, 525 Davey Laboratory, The Pennsylvania State University, University Park, PA 16802, USA\\
}

\date{Accepted XXX. Received YYY; in original form ZZZ}

\pubyear{2021}

\begin{document}
\label{firstpage}
\pagerange{\pageref{firstpage}--\pageref{lastpage}}
\maketitle

% Abstract of the paper
\begin{abstract}
Under the hypothesis that self-reproducing  probes have formed a galactic-scale communication network by direct Gravitationally-Lensed (GL) links between neighboring systems, we identify Wolf 359, the third nearest stellar system, as an excellent target for a search for alien interstellar communication emitted from our Solar System. Indeed, the Earth is a transiting planet as seen from Wolf 359, meaning that our planet could pass in an alien communication beam once per orbit. We present a first  attempt to detect optical messages emitted from the Solar System to this star, based on observations gathered by the TRAPPIST-South and SPECULOOS-South robotic telescopes. While sensitive enough to detect constant emission with emitting power as small as 1W, this search led to a null result. We note that the GL-based interstellar communication method does not necessarily require to emit from the so-called "Solar Gravitational Line" (SGL), starting at $\sim$550 au from the Sun, and that the probe(s) could be located closer to the Sun and off-center relative to the
SGL, at the cost of a smaller but still significant gain compared to a non-GL-boosted communication strategy. Basing on this consideration, we searched in our data for a moving object whose motion would be consistent with the one of the hypothesized alien transmitter, assuming it to use a solar sail to maintain its distance to the Sun. We could not reliably identify any such object up to magnitude $\sim$23.5, which corresponds to an explored zone extending as far as Uranus' orbit (20 au).
\end{abstract}

% Select between one and six entries from the list of approved keywords.
% Don't make up new ones.

%%%%%%%%%%%%%%%%%%%%%%%%%%%%%%%%%%%%%%%%%%%%%%%%%%

%%%%%%%%%%%%%%%%% BODY OF PAPER %%%%%%%%%%%%%%%%%%

\section{Introduction}

\cite{G14} (hereafter G14) hypothesized that if the Milky Way had been colonized/explored by self-reproducing interstellar probes \citep{F80, V80},  they could form an efficient galactic-scale communication network by direct links between neighboring systems, using 
the systems' host stars as Gravitational lenses (GL; \citealt{VE, Maccone}). Under this hypothesis, alien Focal Interstellar Communication Devices (FICDs) should be present in the solar system,   in the focal region of some of the nearest stars, at more than 550 au from the Sun. Basing on this hypothesis, G14 proposed a novel Search for Extra-Terrestrial Intelligence (SETI) \citep{Tarter} strategy consisting in the intense multi-spectral  monitoring of the focal regions of the nearest stars, with the hope to catch a communication leakage  from the FICDs. This strategy has been explored further by subsequent studies \citep{ Gertz18, Gertz21, Hippke20b, Hippke20, Hippke21, Kerby2021, Marcy21}.
 
In this paper, we focus on the possibility to detect the {\it interstellar}  messages from a specific FICD ($vs$ its {\it local} communications with probes in the Solar System, as was recently attempted by \cite{Marcy21}). We identify Wolf 359,   the third nearest stellar system, as the best target for such a search, and we present a first attempt to detect optical messages emitted from the Solar System to this star.

\section{Attempting to detect an interstellar message from an alien FICD: practical considerations}

A first element to consider for the envisioned search is the spectral range. 
Because of the refractive and scattering effects of its plasma, the solar corona acts on photons grazing the Sun as a divergent lens opposing the effect of the solar gravity \citep{VE, TA03, GS11, Hippke21}. 
 This deflection effect is especially strong for radio waves. It can be modeled by the following formula
 relating the divergent deflection angle $\theta_{pl}$ to the impact parameter $b$ (in solar radius $R_\odot$)
 and frequency $\nu$ of the photons:
 \begin{equation}\label{eq:1}
\theta_{pl}(b,\nu) = \bigg(\frac{\nu_0}{\nu}\bigg)^2 \lbrack2952 b^{-16} + 228 b^{ -6} + 1.1 b^{-2} \rbrack \textrm{,} 
\end{equation}\noindent where $\nu_0 =$ 6.32 MHz \citep{TA03}. The actual deflection angle of a light ray grazing the
Sun is the sum of $\theta_{pl}(b,\nu)$ and  the gravitational convergent deflection angle $\theta_{gr}(b)$. 
Assuming that the space-time around the Sun is well described by the Schwarzschild metric, 
$\theta_{gr}(b)$ is given by:
\begin{equation}\label{eq:2}
\theta_{gr}(b) = \frac{4 G M_\odot}{c^2 b R_\odot} \textrm{,}
\end{equation}\noindent where $G$ is the gravitational constant, $c$ is the speed of light, and $M_\odot$ is the 
solar mass \citep{VE}. The opposite directions of the two deflection effects make the lensing impossible for frequencies below a critical value $\nu_{crit}$ that depends on the impact parameter $b$. It can be computed with the following formula \citep{TA03}:
\begin{equation}\label{eq:3}
\nu_{crit}(b) = \bigg\lbrack \nu_0^2 \frac{R_\odot c^2}{4 G M_\odot}  (2952 b^{-15} + 228 b^{ -5} + 1.1 b^{-1} ) \bigg\rbrack^\frac{1}{2}
\end{equation}\noindent This formula gives $\nu_{crit}$ = 64 GHz for $b = 1.1$. Still, in practice, latitudinal dependencies and fluctuations in the coronal electron density will strengthen the divergent effect of the coronal plasma, making necessary to emit at frequencies beyond 1THz (0.3 mm),  to maximize the amplitude of the microlensing effect and to place the FICD at the shortest possible distance from the Sun \citep{TA03}. The microwave range used by classical  SETI experiments ($\sim$1 to 10 GHz) is thus not suited to a search for gravitationally focused extraterrestrial communications. 

A second element to consider is the strong geometrical constraint for this special SETI strategy. To estimate it, it is required to take into account that,  for a given distance to the Sun of the hypothesized FICD, only photons grazing the Sun within an extremely narrow effective communication ring (ECR) around the Sun's disk will be focused to the targeted nearby star \citep[e.g.][]{VE, Maccone}. 
Assuming a 100m-radius receiving antenna around Alpha Centauri A, we performed ray tracing computations based on formula 1 and 2 for UV to IR wavelengths that resulted to widths of a few cms for this ECR. We thus assume here that, for the sake of efficiency, the assumed FICD transmitter is composed of an array of lasers emitting narrow beams centered precisely on the ECR (Fig. 1), equivalent to a single more powerful laser emitting an annular beam. In this hypothesis, detecting an FICD's emission to a nearby star can only be done if the observer is within one of these narrow beams, putting a stringent geometrical constraint on the project concept. For an Earth-based observer, this means that the  Earth's minimum impact parameter has to be close to 1 as seen from the FICD, and thus also from the targeted nearby star. In other words, the Earth has to be a transiting  (or nearly transiting) planet for one of the nearest stars to give this SETI concept a chance of success, so the target star has to be very close to the ecliptic plane. With its nearly circular orbit and its semi-major axis  215 times larger than the solar radius, the Earth has a mean transit probability $<0.5$\% for any random star of the solar neighborhood \citep{Winn}. We can thus consider  ourselves very lucky that the third nearest stellar system, Wolf 359, a very low-mass red dwarf of spectral type M6.0V at a distance of  7.8 light-years \citep{Jenkins09, Sebastian21}, lies at only $0.2^{\circ}$  of the ecliptic plane, in the Leo constellation. For Wolf 359, the Earth has a minimum impact parameter about $215\times\sin{0.2^{\circ}} = 0.75$, i.e. it is a transiting planet \citep{K21}.  Furthermore, Wolf 359 is a single star, i.e. it does not undergo strong dynamical perturbations that would make necessary to change constantly the position of a putative FICD in its vicinity \citep{Kerby2021}. These considerations makes Wolf 359 an excellent target for a first attempt to detect an alien interstellar communication from the Solar System.

\begin{figure}[]
\begin{center}
\includegraphics[width=8cm]{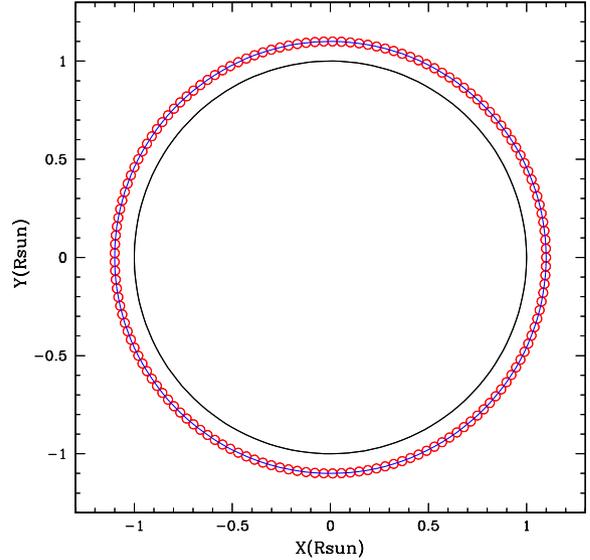}
\caption{Illustration of the Sun (black circle) as seen from the hypothesized FICD. The blue circle represents the ECR, i.e. 
the ring for which photons from the FICD will reach the target receptor. Its actual thickness is a few cms. 
The small red circles represent the laser beams emitted by the FICD in the Sun's plane. 
For this illustration, an impact parameter of 1.1 and laser wavelength and waist of, respectively, 500nm and 1m were assumed. Each laser beam is $\sim$2.5 times wider than the Earth.   }
\end{center}
\end{figure}

A third element to consider is the sensitivity. To estimate it, we assumed the following parameters for the communication  link from the Solar System to Wolf 359:\begin{itemize}
\item A  mean emission wavelength of 500 nm (600 THz) for the lasers of the emitter. A much higher frequency than the lower limit of 1THz mentioned above makes sense, as Wolf 359 is a chromospheric active star \citep{Lin2021} and could thus have a much denser and variable coronal plasma than the Sun;
\item A mass and radius of 0.11 $M_\odot$ and 0.14 $R_\odot$ for Wolf 359 \citep{Sebastian21};
\item An impact parameter $b$ of 1.1 for the light rays grazing both the Sun and Wolf 359. As Wolf 359 is smaller than the Sun, this means that the distances of the first FICD to the Sun and of the second FICD to Wolf 359 have to be tuned to account for the required extra deviation of the light rays (Fig. 2). Our computed distances are 665 au for the emitter to the Sun, and 117 au for the receptor to Wolf 359.
\item For the receptor around Wolf 359, a circular receptor with a radius $R_{rec}$ = 100 m;
\item For each laser, a waist (radius of the beam at its emission) of 1m, corresponding thus to a space telescope of 2m diameter; 
\end{itemize} The Rayleigh length $z_R$ of a laser is the distance from the emission source along the propagation direction corresponding to the doubling  of the area of the beam's cross section. Assuming a Gaussian beam, it is given by : \begin{equation}\label{eq:4}
z_R = \frac{\pi \omega_{0}^{2}}{\lambda}  \textrm{,} 
\end{equation}\noindent 
where $\omega_{0}$ is the laser waist \citep{laserbook}. Under the assumed parameters, $z_R$ is $\sim$ 6300 km $\sim 1  R_\oplus$. 
The radius of the beam at a radial distance $d$ from its source is then given by \citep{laserbook}: \begin{equation}\label{eq:5}\omega(d) = 
\omega_{0} \sqrt{1 + \frac{d^2}{z_R^{2}}}\end{equation}\noindent At 665 au, $\omega(d)$ is = 15800 kms $\sim$ 2.5 $R_\oplus$, i.e. it could
encompass completely the Earth. To cover the whole ring of impact parameter 1.1 surrounding the Sun (Fig. 1), an array of 152 lasers 
would be required. With bigger lasers (and thus smaller Rayleigh lengths), the beams would be narrower, so the global  efficiency and the communication rate 
would be improved, but more lasers would be required to completely surround the Sun, resulting in an increase of the size  but also 
of the mass of the FICD, and thus on the energy required to keep fixed its distance to the Sun. A balance has thus be found between the
 global efficiency of the FICD, its size, and its energetic consumption. In this respect, our assumed value of 1m for the laser waist is 
 nothing more than a  simple work assumption to estimate the potential of detection of the FICD. 
 
 The global efficiency of the FICD is the
  fraction of its emitted photons that will actually reach the receptor. Performing ray-tracing simulations based on Eq. 1 and 2 and the assumptions described above, we derived a width of 22 cm for the ECR. The corresponding fraction can be estimated as the ratio of the effective area of the beam at distance $d$ and its total area: $2 \times  0.22  \times 15.8\times 10^6 / (\pi \times (15.8 \times 10^6)^2) = 8.8 \times 10^{-9}$. The factor 2 in this equation reflects the fact that the intensity in the center of a gaussian beam is about twice larger than the mean intensity in the beam. For comparison, one can compute that the efficiency of the same laser without using the Sun and Wolf 359  as gravitational lenses is $1.5\times10^{-15}$. The resulting gain brought by the gravitational lensing is the ratio of these two efficiencies: $2.9\times 10^6$ = 64.6 dB. This gain does not depend on the number of used lasers, i.e. if the whole ECR is used or not. 
 
\begin{figure*}
\begin{center}
\includegraphics[width=16cm,angle=0]{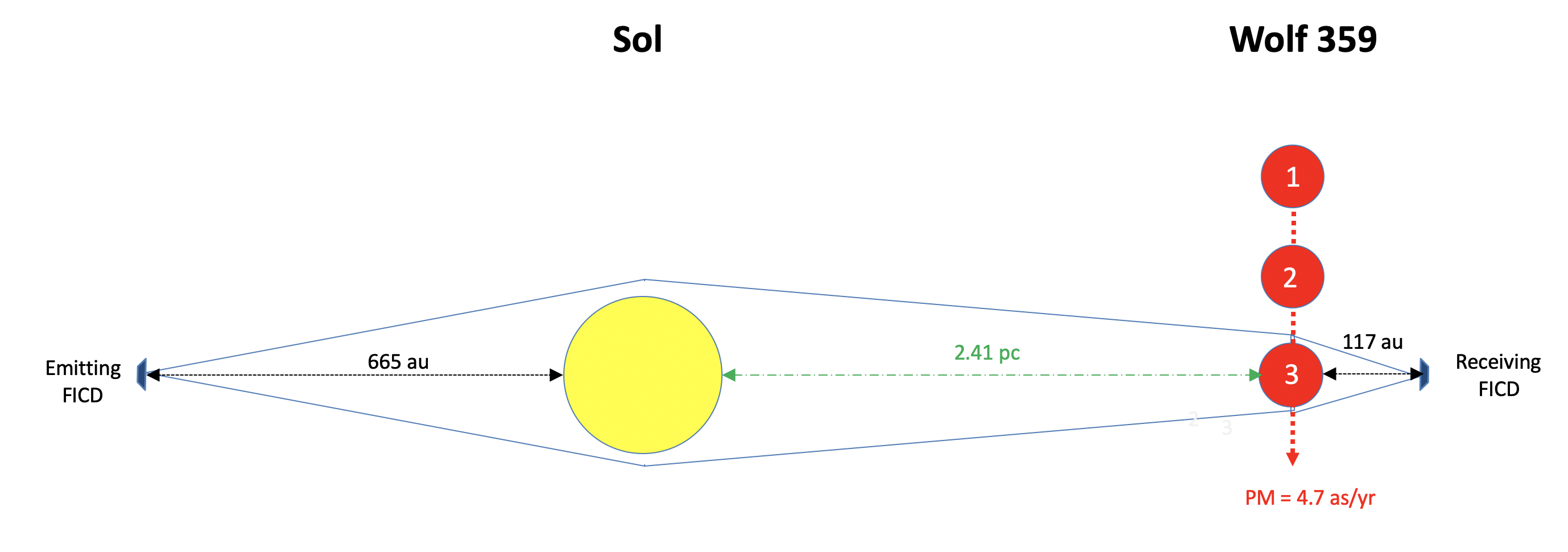}
\caption{Illustration showing the geometry of the hypothesized communication link from the solar system to the Wolf 359 system. The distances and stellar sizes are not to scale. Wolf 359  is shown at 3 different positions. Position 1 corresponds to the time
of the emission of the photons that we receive from it now. Position 2 corresponds to its current position. Position 3 corresponds
to the time it will receive the photons emitted now by the FICD. }
\end{center}
\end{figure*}

At the core of the beam, the average intensity at a distance $d$ from the beam waist  $I_c(d)$ can be estimated as  twice the total emission power $P_0$ divided 
by the area within the radius $\omega(d)$:
\begin{equation}\label{eq:6} 
I_c(d) = 2 \times \frac{P_0}{\pi \omega^2(d)} \end{equation}\noindent 
Assuming an emission at a mean power of 1W per laser (152W in total then for the whole lasers array), $I_c(d)$ is $\sim2.6 \times10^{-15}$ 
W/m$^2$, which corresponds to the intensity of a solar-type star of visual magnitude $\sim$17.6. The firm detection of such a source can be achieved within a few mins of integration with a ground-based telescope of a few dozens of cm aperture 
equipped with a modern CCD camera. Computing the motion of the Earth relative to the Sun as seen from the FICD results in an average duration of 15min for 
the crossing of the laser beam. Assuming exposures of 55s without filter with the robotic 60cm telescope TRAPPIST-South \citep{G11, J11} at ESO La Silla, an 
effective seeing of 2", the target  at airmass 1.5, and the Moon at phase 0.5 and at 50$^{\circ}$ from the target's field of view, we computed a complete noise 
budget (read-out, dark, photon, scintillation, background) that led to an estimated signal-to-noise ratio of $\sim$200 for the detection of the emission after integration over the whole 15min beam crossing. We thus conclude that the hypothesized FICD
 targeting Wolf 359 could be detected with a modest-size optical telescope like TRAPPIST-South, assuming that it emits constantly.  
 
Our ray-tracing computations show that the assumed mean power of 1W per laser for the FICD would translate into $\sim 1.1 \times 10^{10}$ photons reaching the  receptor behind Wolf 359 per second and per laser. Assuming that a  detection sensitivity of 1000 photons per bit is required to ensure a negligible bit-error-rate, the resulting data transfer rate would be $\sim$ 11 Mb per second and per laser, an impressive value for an interstellar communication using such a low-power emitter. In practice, the achieved data transfer rate could be much higher, as sophisticated photon-bit alphabets could achieve negligible bit-error-rate with much lower detection sensitivities (i.e. requiring much less photon per bit). Nevertheless, this estimate has an optimistic side too, as it  neglects the noise brought by the upper atmosphere of Wolf 359 (see Section~\ref{section:my}). 

\section{Attempting to detect an interstellar message from an alien FICD: a first try with TRAPPIST-South and SPECULOOS-South}
 
 Basing on the considerations described above, we performed two searches for a GL-based communication from the Solar System to Wolf 359.
 The first search was carried out in 2015 using the TRAPPIST-South telescope at ESO La Silla Observatory (Chile) \citep{G11, J11}, 
 and the second one in 2018 using  the telescope {\it Europa} of the SPECULOOS-South facility  at ESO Paranal Observatory (Chile) \citep{ Burdanov18, Delrez18, J18, Sebastian21}. Both attempts were done during a transit of the Earth as seen from the putative FICD (i.e. during an occultation of the Earth as seen from Wolf 359). Both were done in the optical range. These observations and their analysis are described below (Sect. 3.2), but first we present the methodology used to compute the astrometric position of the FICD and its position relative to Earth during the observations.
 
 \subsection{Computation of the astrometric position of the FICD and its position relative to Earth}
 
 Using as input the barycentric position (ICRS/J2016 system) and proprer motion in right ascension  (RA) and declination (DEC) of Wolf 359 as measured by Gaia (DR3 release, \citealt{GaiaDR3}), we computed the ICRS coordinates of Wolf 359 at the time of each observation. To the obtained RA and DEC, we then added twice the product of the corresponding proper motion and the distance to Wolf 359 in light-years  (Dly = 7.86 light-years) to compensate for the finite nature of the speed of light. Indeed, in that star's reference system, its astrometric position corresponds to the location  it occupied Dly years ago. Furthermore, as a sniper aiming for a moving target, the FICD has to aim for the position that Wolf 359 will occupy Dly years from now (Fig. 2). Given the high proper motion of Wolf 359, the amplitude of this correction is significant:  -59.9 arcsec and -42.4 arcsec in RA and DEC, respectively. Once this correction done, the barycentric equatorial coordinates of the FICD were computed simply by removing/adding 180 deg from the obtained RA and DEC. We did not consider aberration, as its effect is taken into account in the pointing model of the telescopes. 

We then computed the FICD coordinates as seen from Earth and not from the Sun. First, a celestial transformation to the ecliptic system was applied to the barycentric RA and DEC. The ecliptic coordinates of the Sun at the time of each observation were then computed, providing all the inputs required to move to a geocentric system. A reverse transformation to the equatorial system led finally to the RA and DEC of the FICD at the time of each observation. For each observation, we also used the ecliptic positions of the FICD and the Sun to compute the distance between the center of the Earth and the center of the Sun on the plane of the sky as seen from the FICD (see Fig. 3 and 5), so to predict the enter and exit of the Earth into the FICD beam under an assumed communication strategy (impact parameter, wavelength, laser waist).
   
 \subsection{TRAPPIST-South observations}
 
 TRAPPIST-South is a 60cm robotic telescope located at La Silla Observatory (Chile) and dedicated to the observation of exoplanet transits and small bodies of the solar system \citep{G11, J11}.  It has a Ritchey-Chretien design, and is coupled to a German equatorial mount. It is equipped with a thermoelectrically-cooled 2k $\times$ 2k CCD 
camera with a field of view of 22' $\times$ 22' (pixel scale = 0.65") and a quantum efficiency larger than 50\% from 300 to 920nm. Pointing on the ICRS coordinates RA= 22h56m20.81s and DEC= -06d59m28.4s (computed as described in Sec. 3.1), TRAPPIST-South observed during the whole night of 4$^{th}$ to 5$^{th}$ Sep 2015, from 23h24 UT to 10h13 UT, which corresponded to the start of the transit of the Earth as seen from the putative FICD (Fig. 3). To ease the confirmation of a putative extra-source in the images of the night, the same field was observed again on 5 and 8 Sep for a few hours. All these observations were carried out without any filter, so to provide a spectral coverage as wide as possible (from $\sim$300 to $950$nm). An exposure time of 55s was used for all the taken images. 536 images were taken on 4 Sep 2015, and 140 and 152 on 5 and 8 Sep, respectively.

For the three runs, a standard pre-reduction (bias, dark, flat-field correction) was applied to the images, then they were aligned and combined. The resulting high signal-to-noise stack image was then plate-solved, and the expected position of the FICD was over-plotted (Fig. 3). Comparing the stacked image of the first night to the ones of the second and third night, a visual search for an extra-source in the expected position range of the FICD was performed. No extra source could be identified. 

Figure 4 shows the evolution of the Earth-Sun distance on the plane of the sky during the night of 4 Sep 2015 as seen from the putative FICD. 
Assuming a focal impact parameter of 1.1, i.e. to an ECR with a radius of 1.1 Sun radius (corresponding to a FICD-Sun distance of 663 au, see above), the Earth should have crossed this ring around  2457270.82 JD = 7h40 UT. Under our assumptions on the laser waist described in the previous section, and taking into account the Earth orbital speed of $\sim$30km/s, the telescope should have remained in the laser beam for $\sim$8 min. Still, this simple estimate neglects the fact that, as seen from the FICD, the Earth's transit has a high impact parameter (0.73), and so the Earth's path is nearly tangential to the focal annulus, which results in a much longer time within the beam of $\sim$25 min. Basing on this estimate, we combined the images of the night ten by ten, and inspected the resulting stacked images in search of an extra source appearing at the expected position range of the FICD  for one or a few of them. We could not detect any transient extra-source, but noticed that a very faint source  (estimated $g$-magnitude $\sim$ 22) was located  at the ICRS coordinates RA 22:56:20.77 DEC -06:59:28.4, which corresponds approximately to the position expected for the FICD during the Earth's transit (Fig. 4). Nevertheless, this source is also present in the stacked images of the two other TRAPPIST-South runs, while the FICD should have been at other positions because of its large parallactic motion (see Fig. 3). Furthermore, if this source had been the searched FICD, it should have drifted significantly  on the sky during the night of 4-5 Sep 2015 (see Fig. 3),  again because of its large parallax, which was not the case.  

To ensure that this faint source could not have hidden the emission of the hypothetical FICD, we studied its photometric stability. For that purpose, we used IRAF/DAOPHOT \citep{Stetson} to extract the fluxes within a circular aperture of 6 pixels radius  (4") centered  on its position in the stacked image of the night. As this star is too faint to be visible in individual images, we fixed the position of the center of the aperture. We measured the background in an annulus extending from 18 to 28 pixels of the center of the aperture. Figure 5 shows the resulting light curve. In addition to an increase of the scatter at the beginning and at the end of the night, one can notice several outliers with a significant flux excess. Examining the individual images and a movie of them, we noticed that the first group of outliers corresponds to the passage of an asteroid ($g \sim 19.5$) right in the aperture . The other outlier corresponds to a cosmic hit. 

In a last step, we used the mean  $g$ magnitude of the faintest stars registered in the Gaia DR3 catalogue visible in the images stacked ten by ten to estimate the highest visual magnitude for which a source would have been detected by visual inspection. We reached a value of 21.8, much larger than the magnitude of 18.3 that we estimated under the assumptions described in Section 2. This magnitude corresponds to the one of the source at RA 22:56:20.77 DEC -06:59:28.4. This leads us to the conclusion that our observations had the sensitivity to detect the communication of the putative FICD to Wolf 359 under the assumptions described in Section 2.

\begin{figure}
\begin{center}
\includegraphics[width=6cm]{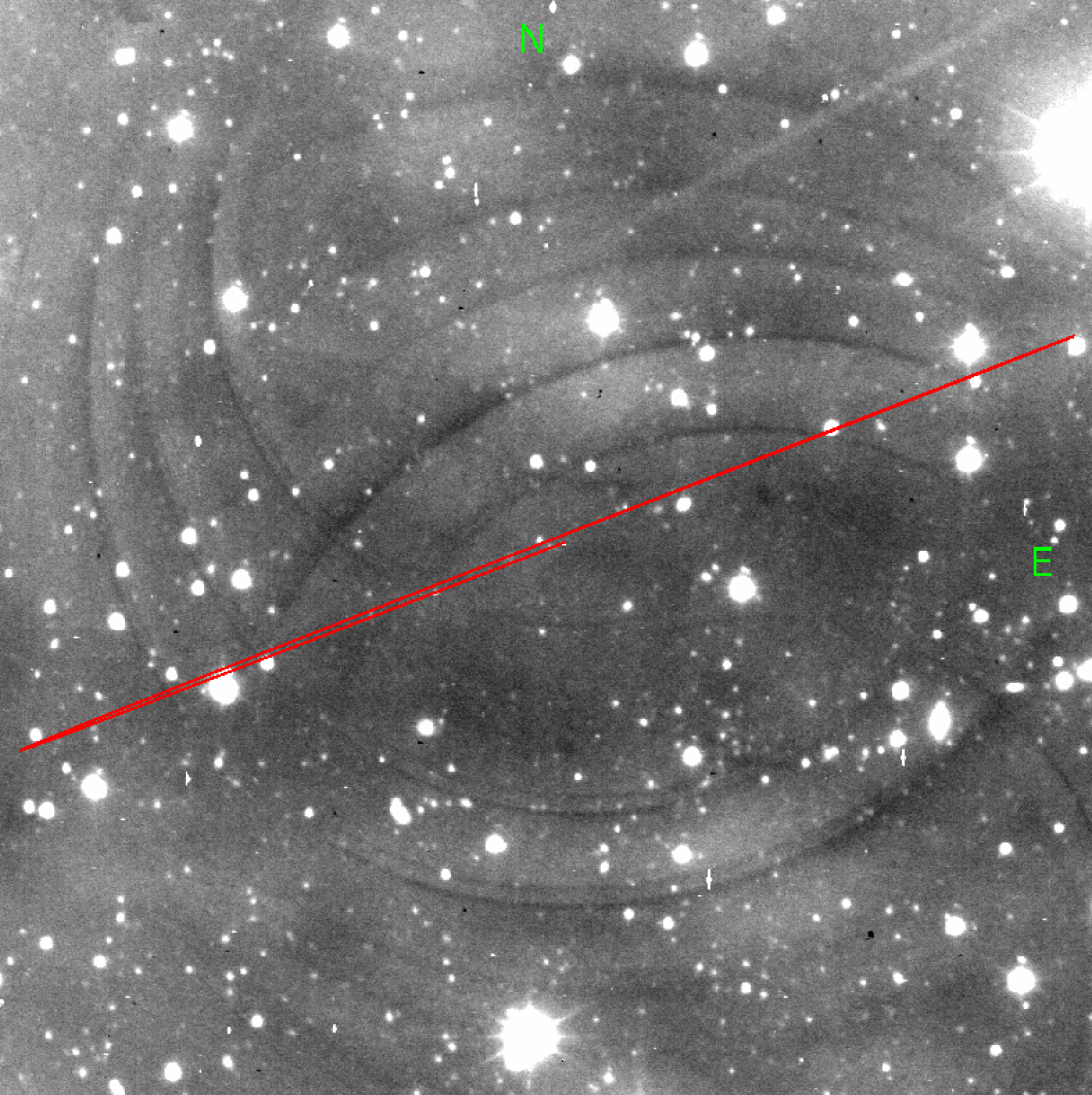}
\includegraphics[width=6cm]{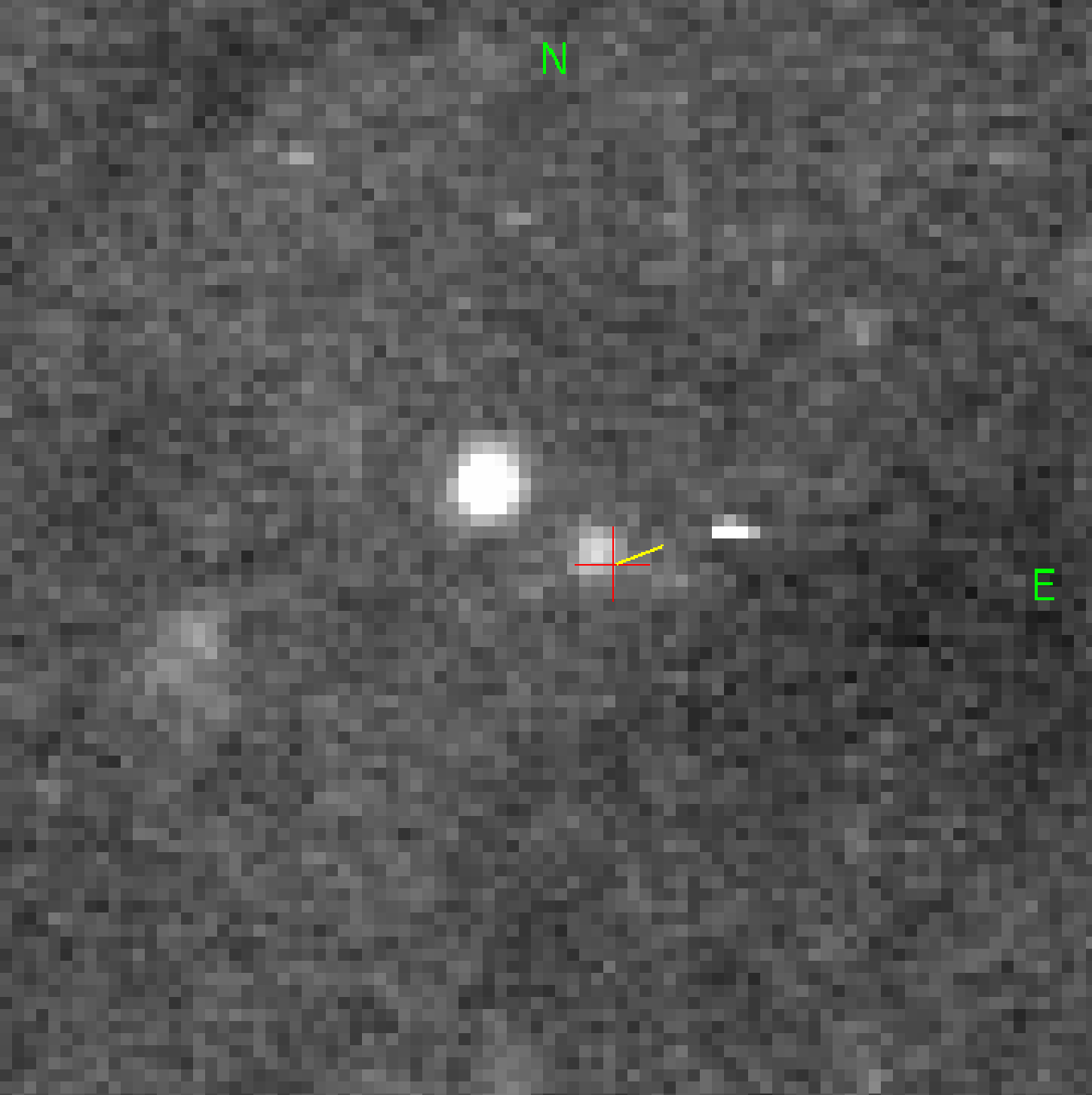}
\caption{$Left$: 10 by 10 arcmin crop of the center of the stack (mean) of all images taken by TRAPPIST-South during the night of 4-5 Sep 2015.  The position of the hypothetical FICD (under the assumptions of Section 2)  between 26 Aug 2015 to 26 Aug 2016 is shown in red. This large motion on the sky is due the parallax effect combined to the compensation for the proper motion of Wolf 359.  $Right$: 10 by 10 arcsec zoom on the path of the hypothetical FICD during the observations of TRAPPIST-South on 4-5 Sep 2015 (yellow line). The red cross shows the position that should have occupied the FICD at the inferior conjunction of the Earth (i.e. in the middle of our planet's transit as seen from the FICD). The bright spot at the east of the FICD location is a hot pixels cluster.}
\end{center}
\end{figure}

\begin{figure}
\begin{center}
\includegraphics[width=8cm]{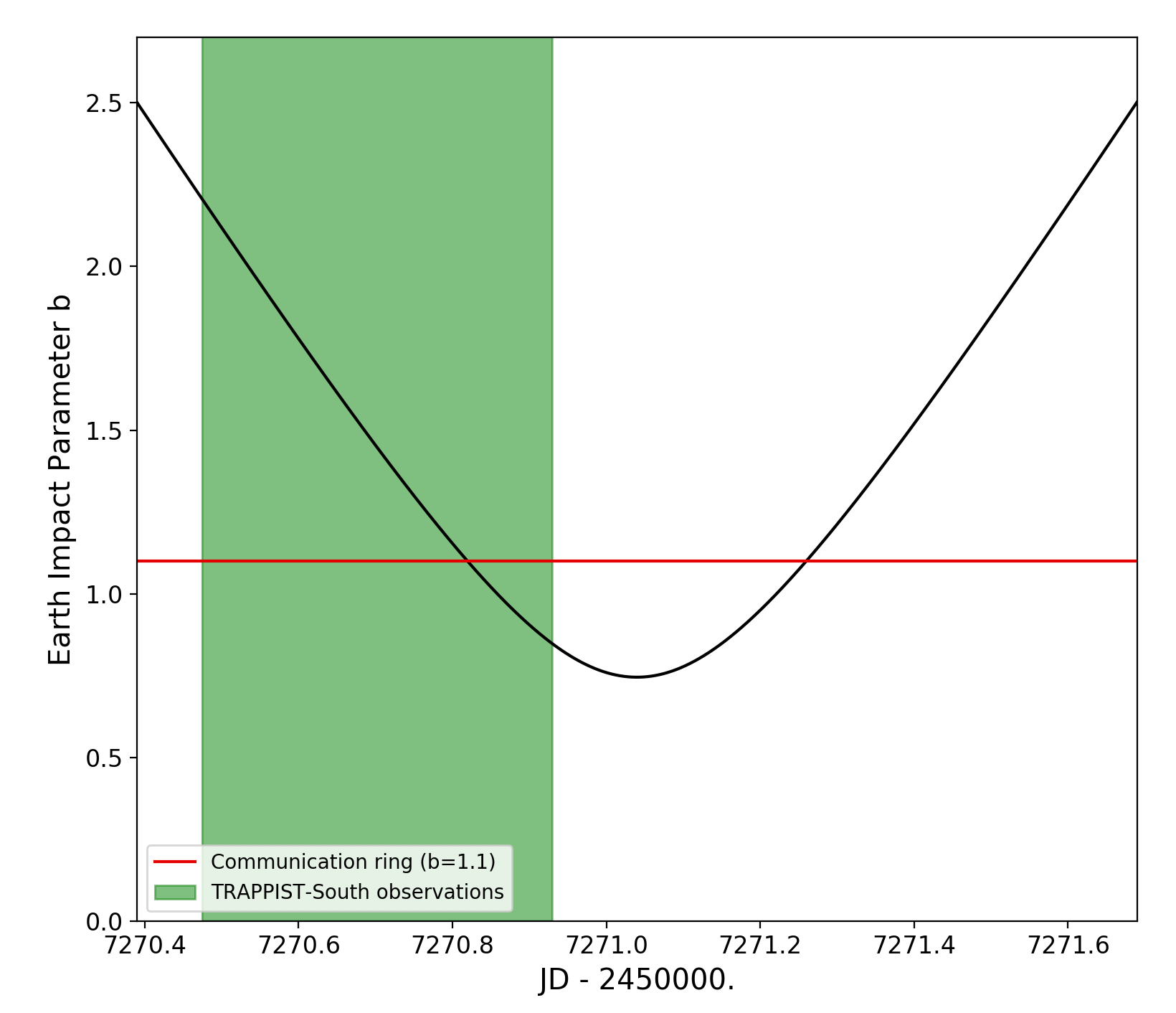}
\caption{Evolution of the impact parameter of the Earth as seen from the SGL of Wolf 359 around the TRAPPIST-South observations of 2015. The red line corresponds to an impact parameter of 1.1. The green zone corresponds to the TRAPPIST-South observations. The intersection of the red and black lines correspond to the crossing of the putative FICD's beam by the Earth. }
\end{center}
\end{figure}

\begin{figure}
\begin{center}
\includegraphics[width=8cm]{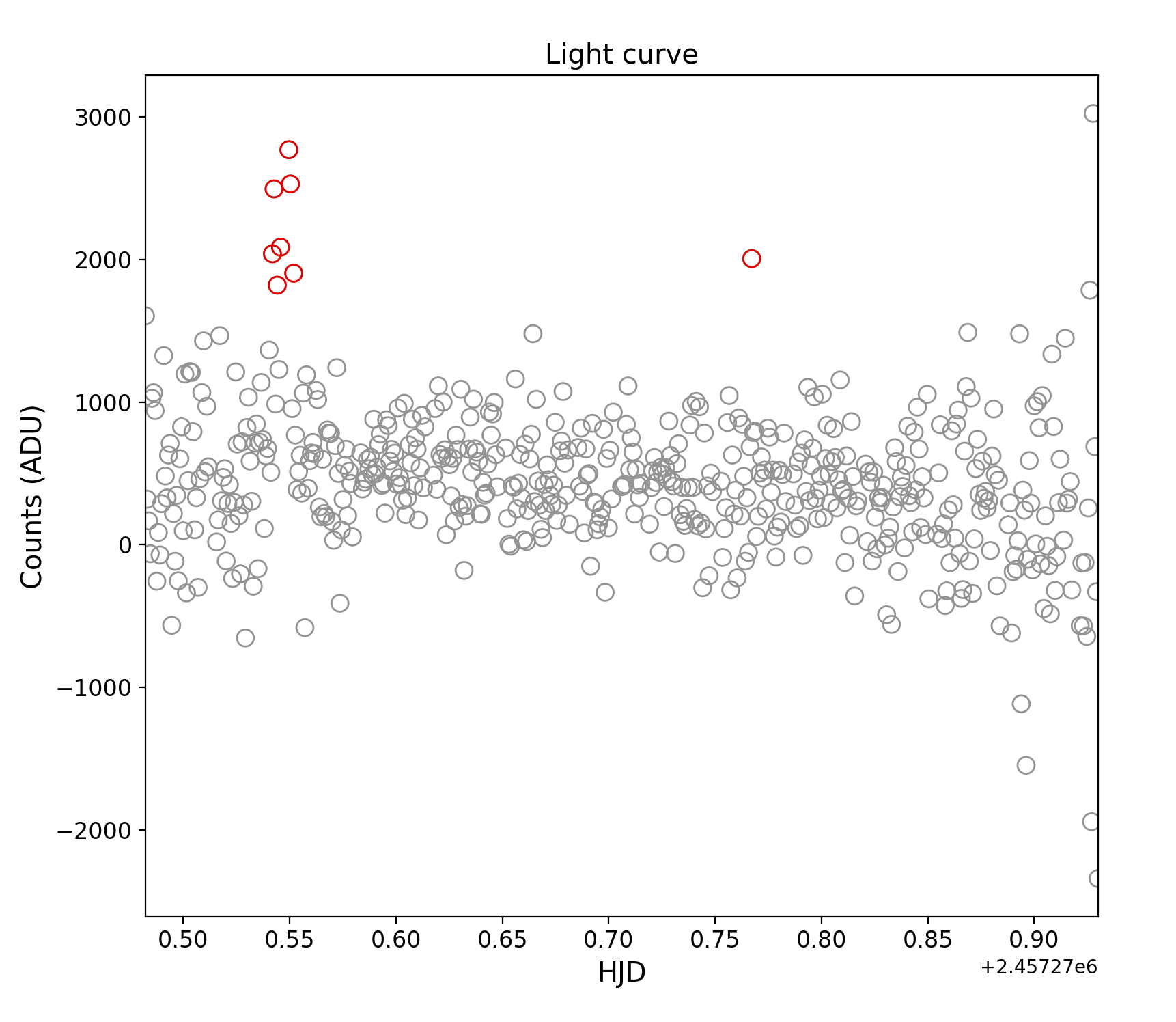}
\caption{Sky-subtracted light curve of the source located approximately at the position of the hypothetical FICD in the TRAPPIST-South images. The points in red have a significant  flux excess. The first group of red points is due to the passage of a $g \sim 19.5$ asteroid in the aperture. We attributed the other outlier to a cosmic hit. }
\end{center}
\end{figure}

 \subsection{SPECULOOS-South observations}
 
 SPECULOOS-South is a facility composed of four 1m robotic telescopes located at Paranal Observatory, Chile \citep{Burdanov18, Delrez18, J18}. It is the core facility of the SPECULOOS project that aims to explore all nearby ultracool dwarf stars (spectral type later than M6) within 40 pc for transiting planets \citep{G18, Sebastian21}.  Each of its four telescopes has a Ritchey-Chretien design and is mounted on an equatorial mount with a special design making unnecessary to flip the telescope position at meridian. All telescopes are equipped with a thermoelectrically-cooled deep-depletion 2k$\times$2k camera  with a field of view of 12' $\times$ 12' (pixel scale = 0.35") and a quantum efficiency larger than 50\% from 420 to 950nm. The SPECULOOS-South telescope $Europa$ observed the whole night of 4$^{th}$ to 5$^{th}$ Sep 2019, from 0h UT to 9h24 UT, which corresponded to the start of the transit of the Earth as seen from the putative FICD. All these observations were carried out without any filter, so to provide a spectral coverage as wide as possible. An exposure time of 50s was used for the 492  taken images. 
 
 The same reduction and analysis procedures as for the TRAPPIST-South data were applied. During the reduction, we noticed that the tracking of the telescope performed poorly during the night, because of the use of a bad pointing model (at that time, the telescope was still in commissioning phase). Because of this bad tracking resulting in a shift of  the stars of $\sim$1" from one image to the other, we had to discard all images taken after 2h55 UT, keeping only the first 150 images. Fortunately, the crossing of the ECR by the Earth was supposed to take place within the first 3hrs of the night, with a start of the transit of the Earth as seen from the FICD computed to happen around 1h45 UT (Fig. 6). 
 
As for TRAPPIST-South, we examined the individual images and the images stacked ten by ten. We did not find an extra-source at the expected position of the FCID in any of those images (Fig. 6 and 7 are the SPECULOOS counterparts of Fig. 3 and 4).
 
Using the same method than for the TRAPPIST-South observations, we estimated the highest $g$-band magnitude for which a source would have been detected by visual inspection of the images stacked ten by ten. We reached a value of 22.5, high enough to ensure that the putative FICD would have been detected under the assumptions described in Section 2. 

\begin{figure}
\begin{center}
\includegraphics[width=6cm]{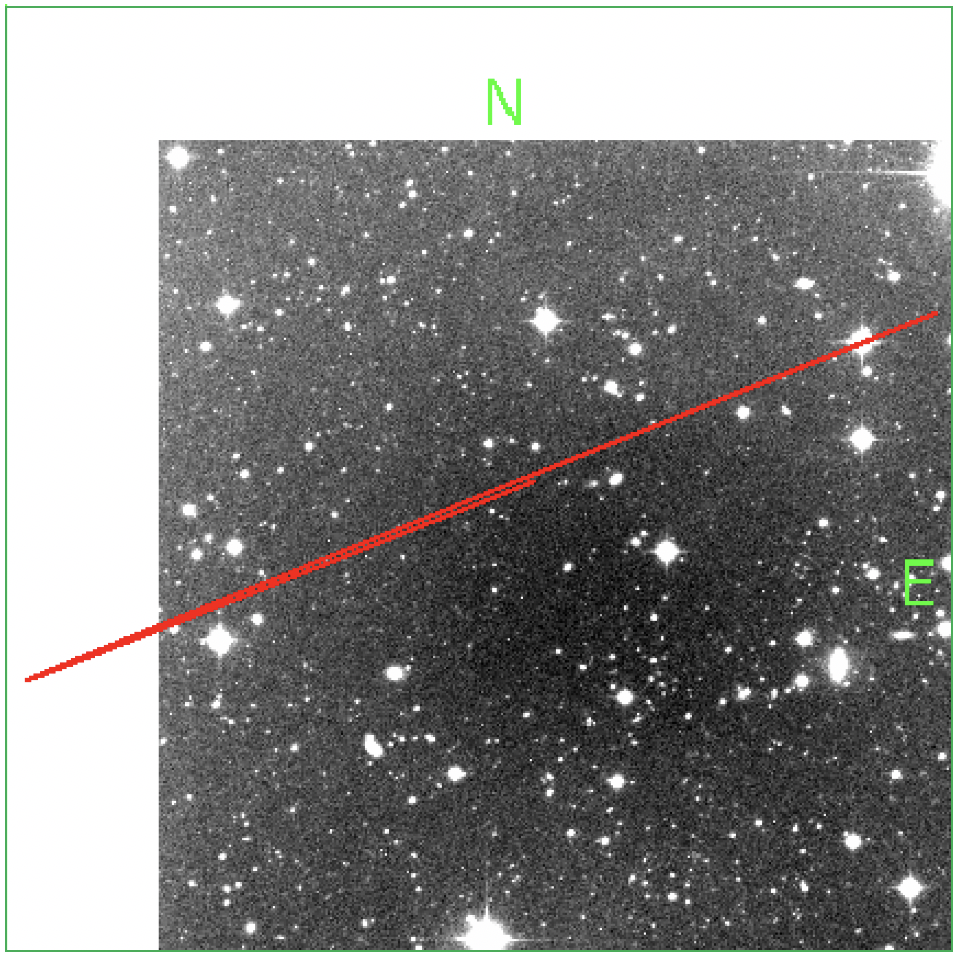}
\includegraphics[width=6cm]{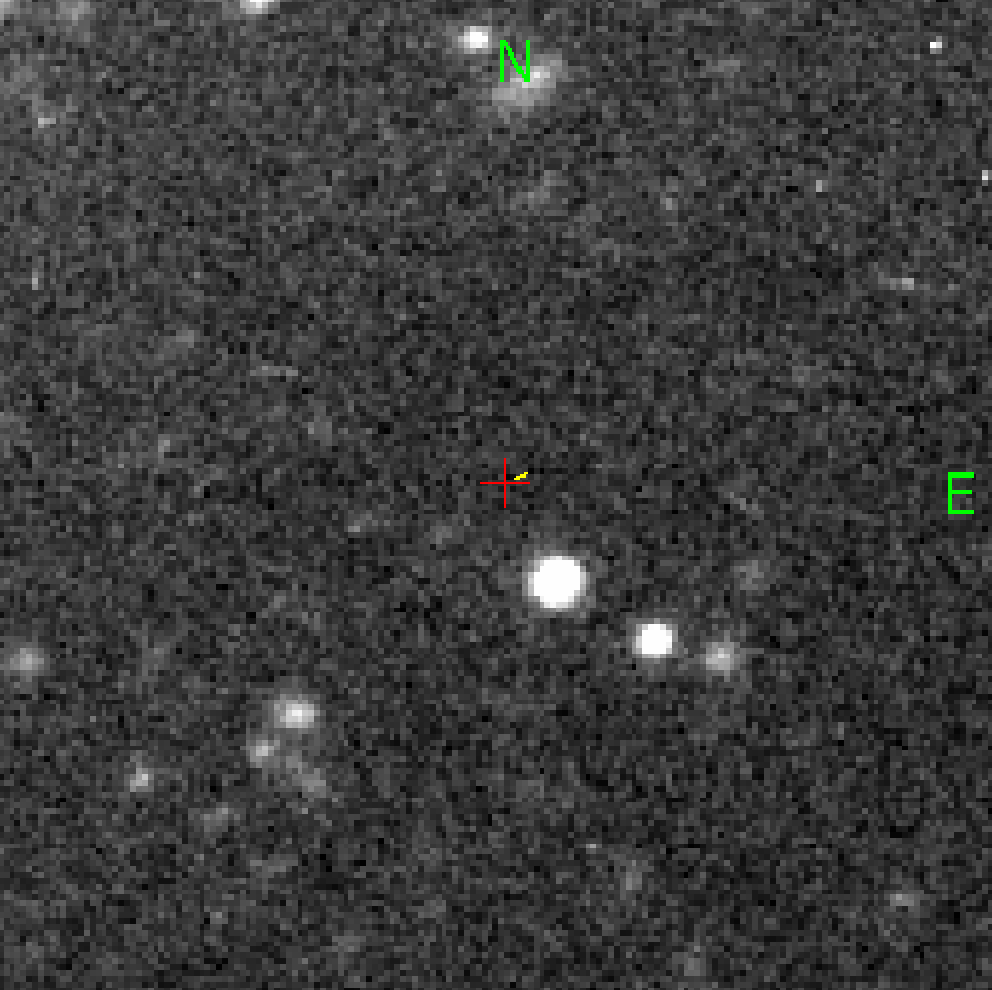}
\caption{Same as Fig.~3 but for the 150 first images taken by SPECULOOS-South/$Europa$ of 4-5 Sep 2018. The whole stack of the images ($left$) is smaller than 10 $\times$ 10 arcmin because of the shift of the images due to the tracking problem  experienced by the telescope during the night. }
\end{center}
\end{figure}

\begin{figure}
\begin{center}
\includegraphics[width=8cm]{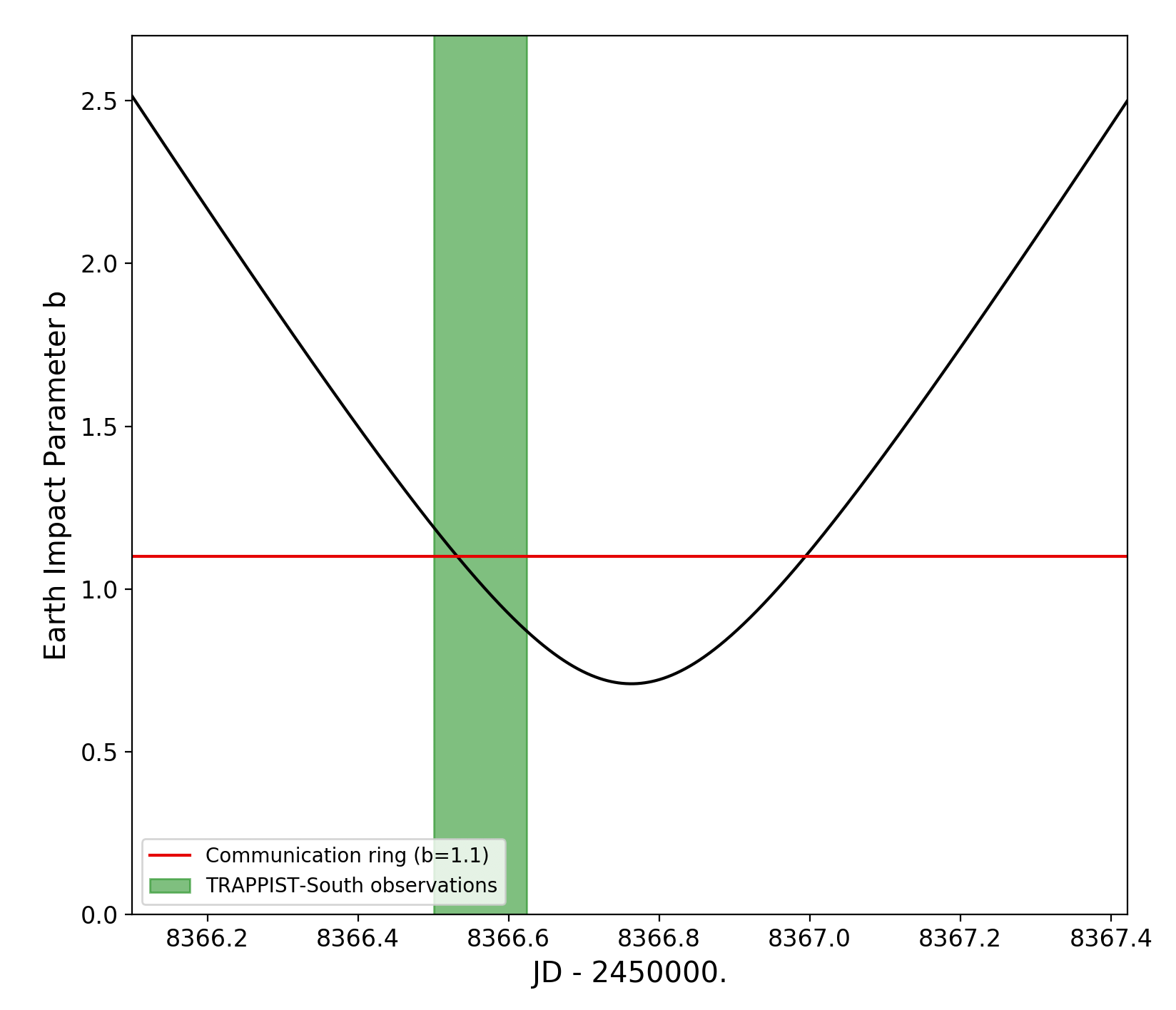}
\caption{Evolution of the impact parameter of the Earth as seen from the hypothetical FICD around the SPECULOOS-South observations of 2018. The red line corresponds to an impact parameter of 1.1. The green zone corresponds to the SPECULOOS-South observations.  }
\end{center}
\end{figure}

 \section{Discussion} 
 \label{section:my}
 
This search for an interstellar optical communication from the solar system to Wolf 359 did not lead  to any  detection. This null result could be explained by a large collection of hypotheses, for instance: our Solar System had never been reached by the probes from an alien technological civilization; such probes are well in our Solar System, but they do not use a GL-based communication strategy; they do use this strategy, but they did not establish a communication link between the solar and the Wolf 359 system; such a link exists, but the corresponding alien FICD did not emit during our observations; it did, but always avoid one of its beams passing on the Earth to remain hidden to the local young technological civilization; it does not use optical wavelengths for its message to Wolf 359; etc.

 Assuming the searched alien FICD does exit and emits continuously, one could wonder if the assumptions presented in Section 2 represent the optimal strategy for such interstellar communication, notably in terms of the spectral range of the emission. The GL-based interstellar communication method makes necessary for the receptor to face its host star, requiring a device to block its light (coronograph or external occulter) while enabling the signal coming from the star's communication ring (assumed here to be at $b$ = 1.1) to reach it \citep[e.g.][]{Hippke18AA}. In the specific case of Wolf 359, the star is a M6.5-type red dwarf with an effective temperature of 2800 $\pm$ 100K, and its spectral energy distribution peaks around 1.1 microns \citep{Pavlenlo06}. Furthermore, it is labeled as a flare star by SIMBAD\footnote{http://simbad.u-strasbg.fr/simbad/sim-id?Ident=Wolf+359}; and it has been observed to show a strong chromospheric and coronal activity \citep{Fuhrmeister04, Schmitt04, Lin2021}. Even if  not a perfect twin, the well-studied star TRAPPIST-1 is probably a good analog to Wolf 359, as it is also a flare star of $\sim$ 0.09 $M_\odot$ with a significant chromospheric activity \citep[e.g.][]{Bourrier17}. A semi-empirical model of its spectral energy distribution is presented in Fig. 8, taken from \cite{Wilson21}.  It shows that, in addition to its photospheric emission peaking at 1.1 micron, the star has a significant emission in X-ray and EUV. This high-energy emission is mostly of chromospheric origin and is variable \citep{Bourrier17, Wilson21}. The same statement should also apply to Wolf 359, as it is an even more active and variable star than TRAPPIST-1. Interestingly, Fig. 8 shows that at the junction of the `chromospheric' and `photospheric' spectra lies a spectral zone of minimal emission, from $\sim$150 to $\sim$250 nm (far and mid-UV). While the very low emission of late-type M-dwarfs in this spectral range could be an issue for prebiotic chemistry on habitable planets \citep{Rimmer18}, it could represents a nice spectral `sweet spot' for a GL-based communication to a late M-dwarf like Wolf 359 or TRAPPIST-1. Another advantage of using this wavelength range instead of the optical range is the improved emission rate, thanks to the narrower laser beams (eq. 4).  These considerations suggest that the spectral ranges 300-920nm and 400-950nm probed by the TRAPPIST-South and SPECULOOS-South observations could not correspond to the optimal spectral range for a GL-based communication from the Solar System to Wolf 359, and that it  could be desirable to reproduce our experiment in the 150-250 nm spectral range, which would require the use of a space-based or a balloon-based telescope. 

\begin{figure*}
\begin{center}
\includegraphics[width=14cm]{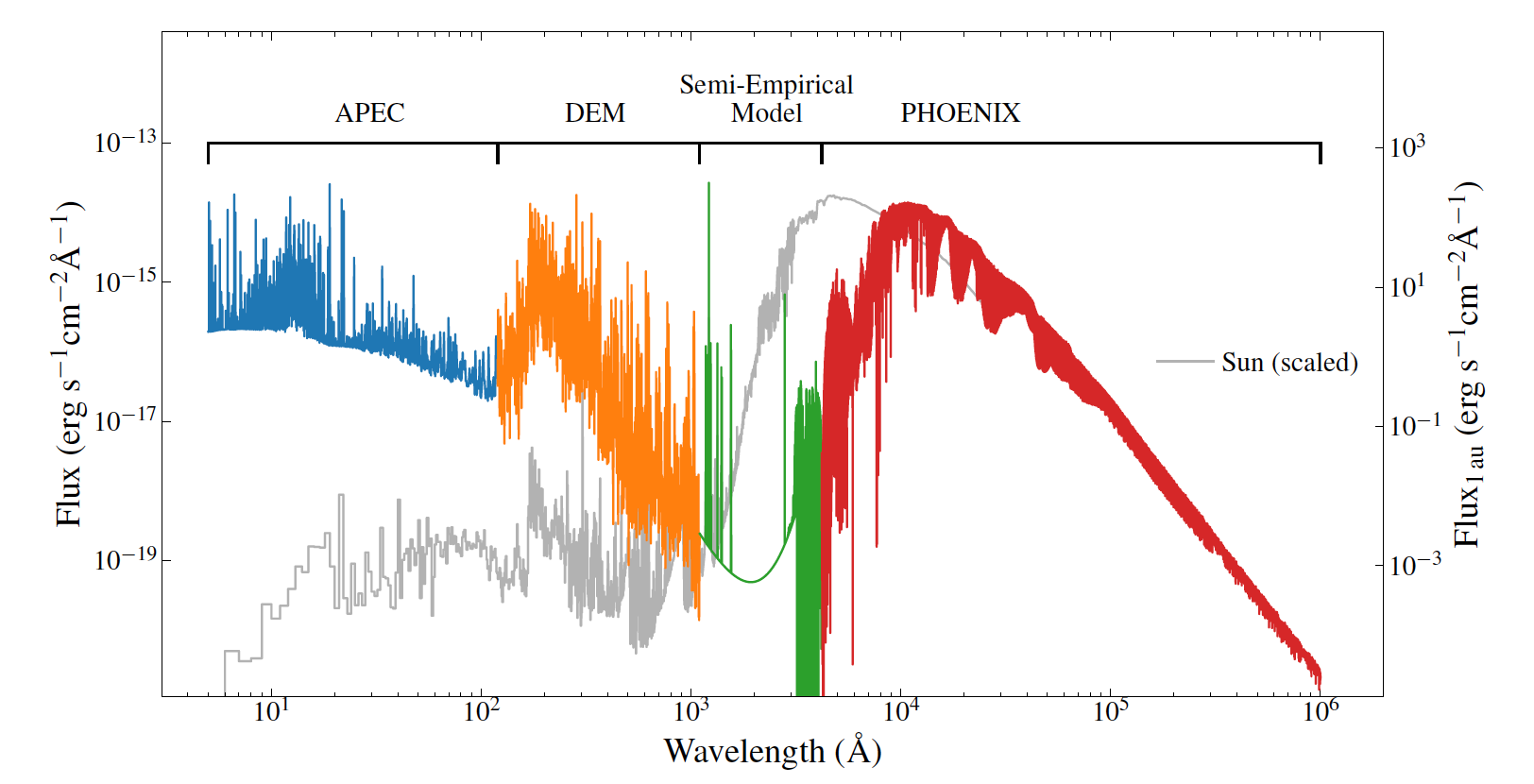}
\caption{Semi-empirical model for the spectral energy distribution of TRAPPIST-1 \citep{Wilson21}.}
\end{center}
\end{figure*}

Another element to consider is the radial position of the emitting probe relative to the Sun.
In Sec. 2, we made the assumption that it was composed of an array of
 1m-waist lasers located in-focus at a computed
distance to the Sun of 665 au. Nevertheless, nothing prevents these
emitters to be separated and independent, and off-center relative to
the focal line (Fig. 9). In such a case, they could be much closer to the
Sun, which would bring two important advantages: their beams in the Sun's plane 
would be significantly narrower, and they would receive
much more energy from the Sun. Nevertheless, there would be a price to pay.
Indeed, the light rays emitted by the laser would reach the Sun's plane
with a different angle, except for those at the exact center of the beam.
Performing new ray-tracing computations confirmed a decrease of the gain 
for larger distances to the focus (Fig. 9). Still, the gain remains
significant up to very small distances to the Sun, and it is more than compensated
by the increase of the solar irradiation with smaller distance. For instance, 
putting the emitter at the distance of 10 au makes the ECR's width (and thus the gain)
shrink by a factor 33, while the stellar irradiation of the FICD would increase by a factor 4400. A point to consider too is that parking the FICD closer to the Sun would also lead to stronger gravitational perturbations from the planets that would require more axial position corrections, and thus more energy, to keep the FICD-Sun-target alignment \citep{Kerby2021}. A possibly interesting option could be to place the FICD close to the outer edge of the Kuiper belt, to avoid too strong gravitational perturbations from giant planets while remaining relatively close of a source of raw material (which could be necessary to maintain the emitter in operational state over thousands if not millions of years) and benefiting from a solar irradiation still $\sim$100 times larger than at the SGL. 

\begin{figure}
\begin{center}
\includegraphics[width=8cm]{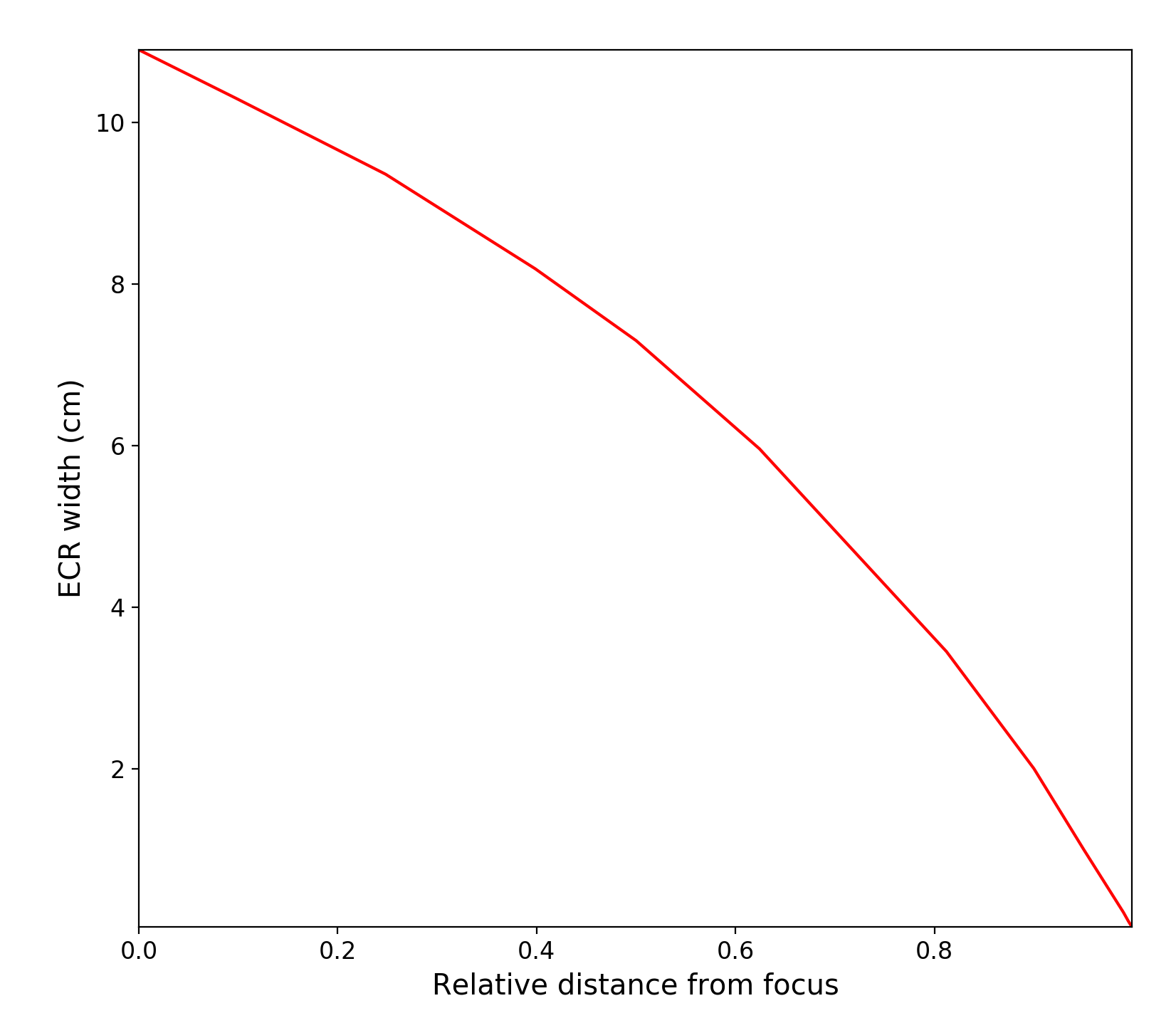}
\caption{Width of the Effective Communication Ring (ECR) around the Sun as a function of the relative shift towards the Sun (0= at focus, 1= at the Sun), as computed from ray-tracing simulations under the assumptions described in the text. }
\end{center}
\end{figure}

Basing on these considerations, we performed new ray-tracing computations, this time assuming a laser emitting with a power of 1W  at 200 nm with a waist of 1 m at a distance of 66.5 au from the Sun. They led to 2e9 photons reaching the 100m-radius receptor opposite to Wolf 359 per second, which corresponds to a gain of 2.1e5 (53 dB) and to an information rate of 2 Mb/s, i.e. 5 times lower than assuming an in-focus laser operating at 500 nm. This information rate could be considered as small, but again it was computed assuming a value of 1000 photons-per-bit for the detection sensitivity. As outlined by \cite{Hippke18b}, this high value could be over-pessimistic. Most importantly, this decrease by a factor 5 of the information rate relative to an in-focus position could be more than balanced by the 100 times largest irradiation of the FICD. 

This alternative hypothesis assuming a FICD composed of one to several off-center lasers located closer to the Sun makes less appealing the search strategy used in this work, as it would be highly unlikely that the Earth would pass in the narrow communication beam(s) of one of these few probes. Nevertheless, their closer distance to the Sun could make possible
their direct detection in reflected light. G14 computed a value of 30.5
for the optical magnitude of a FICD with a payload of 1 ton and a sail
surface density of 0.5g/m$^2$ located at 550 au. This estimate was an
absolute lower limit as it assumed the sail to be seen with a zero phase
angle (i.e. from the Earth transiting the Sun as seen from the FICD),
a Bond Albedo of 1, and a purely specular reflection. Assuming now
an heliocentric distance of 50 au, this estimated
magnitude would shrink to a value of 25.3, which is still out of reach
of a 1m-class telescope like those used in this work, but well within
reach of the biggest existing telescopes. For instance, we used the
online Exposure Time Calculator of the FORS2 instrument of the
ESO Very Large Telescope \footnote{https://www.eso.org/observing/etc/bin/simu/fors\_ima} 
to compute a signal-to-noise of 12 on a
V=25.4 target, assuming a 1hr exposure in Bessel R filter and average
observing conditions (airmass 1.5, Moon illumination 0.5). In reality,
a solar sail would not have a perfectly specular reflection (even if built
by more technology-advanced aliens), so its actual magnitude should
be a bit higher, but it could be compensated by observing without
any filter, using a bigger telescope like the upcoming E-ELT, using
adaptive optics to boost the contrast with the sky background, etc.

It should also be noted that assuming such an off-center architecture
for the FICD increases significantly the uncertainty on the
possible positions of its laser+sail components. Assuming an impact
parameter of 1.1, a central transit of the Earth as seen from the GL
focal point, and a distance to the Sun of 50 au for its components,
the surface to explore would not be a point anymore but a circle of
radius of $\sim$0.65 arcmin = 39". Furthermore, 50 au for the distance is
just an arbitrarily value. The laser(s) could be further away, but they
could also be much closer to the Sun, making necessary to explore
a much larger field of view (but easing the detection in terms of
signal-to-noise). 

While the detailed study of the potential for the direct detection
of such FICD composed of off-center laser+sail units is beyond the scope of this
paper, we performed a first search based on this
hypothesis in our data. This search for faint moving targets was done
using a synthetic tracking (or "shift-and-add") algorithm, a powerful
and computationally expensive method for searching for moving
objects in astronomical images by stacking the images according to
all the possible movements of the object (see e.g., \citealt{1995ApJ...455..342C, 2014ApJ...782....1S}). To perform this search, we used the Tycho
Tracker\footnote{https://www.tycho-tracker.com} software, restricting the search to objects moving slower than 0.5 arcsec/min. Upon completion of the search, we could not
reliably identify any objects which move slower than 0.5 arcsec/min
and which are brighter than our estimated sensitivity upper limit of
$g$-magnitude $\sim$23.5. This limit would correspond to a 1 ton (payload)
FICD located at $\sim$20 au, close to the orbit of Uranus.

Unlike our attempt presented here to detect the communication from a putative
alien FICD to Wolf 359, the search for the reflected light signal from off-axis FICD
does not necessarily require the FICD (and thus its target star) to lie in the ecliptic.
Based on this fact, we have decided to perform a search sensitive to objects up to 
magnitude $\sim$ 26 for the 10-20 nearest stars. This search is ongoing, and its results will
be presented in a forthcoming paper.

\section*{Acknowledgements}

TRAPPIST-South is funded by the Belgian Fund for Scientific Research (Fond National de la Recherche Scientifique, FNRS) under the grant FRFC 2.5.594.09.F, with the participation of the Swiss National Science Fundation (SNF). SPECULOOS-South has received funding from the European Research Council under the European Union's Seventh Framework Programme (FP/2007-2013) (grant Agreement n$^\circ$ 336480/SPECULOOS) and  under the 2020 research and innovation programme (grant agreement n$^\circ$ 803193/BEBOP), from the Balzan Prize Foundation, from the Belgian Scientific Research Foundation (F.R.S.-FNRS; grant n$^\circ$ T.0109.20), from the University of Liege, from the ARC grant for Concerted Research Actions financed by the Wallonia-Brussels Federation, from the Simons Foundation (PI D. Queloz, grant number 327127), from the MERAC foundation, and from the Science and Technology Facilities Council (STFC; grant n$^\circ$ ST/S00193X/1).  M.G. is FNRS Senior Research Associate. 

%%%%%%%%%%%%%%%%%%%%%%%%%%%%%%%%%%%%%%%%%%%%%%%%%%
\section*{Data Availability}

All the data and codes used in this work can be obtained via a request to the first author.

%%%%%%%%%%%%%%%%%%%% REFERENCES %%%%%%%%%%%%%%%%%%

% The best way to enter references is to use BibTeX:

\bibliographystyle{mnras}
\bibliography{GL_Wolf359} % if your bibtex file is called example.bib

% Alternatively you could enter them by hand, like this:
% This method is tedious and prone to error if you have lots of references
%\begin{thebibliography}{99}
%\bibitem[\protect\citeauthoryear{Author}{2012}]{Author2012}
%Author A.~N., 2013, Journal of Improbable Astronomy, 1, 1
%\bibitem[\protect\citeauthoryear{Others}{2013}]{Others2013}
%Others S., 2012, Journal of Interesting Stuff, 17, 198
%\end{thebibliography}

%%%%%%%%%%%%%%%%%%%%%%%%%%%%%%%%%%%%%%%%%%%%%%%%%%

%%%%%%%%%%%%%%%%% APPENDICES %%%%%%%%%%%%%%%%%%%%%

%\appendix

%\section{Some extra material}

%If you want to present additional material which would interrupt the flow of the main paper,
%it can be placed in an Appendix which appears after the list of references.

%%%%%%%%%%%%%%%%%%%%%%%%%%%%%%%%%%%%%%%%%%%%%%%%%%

% Don't change these lines
\bsp	% typesetting comment
\label{lastpage}
\end{document}